\documentclass[runningheads]{svmult}
\usepackage{epsfig}
\usepackage{subeqnar}  
\usepackage{multicol}  
\usepackage{physprbb}  
\makeindex             

\newcommand{\beq}{\begin{equation}}
\newcommand{\eeq}{\end{equation}}
\newcommand{\bea}{\begin{eqnarray}}
\newcommand{\eea}{\end{eqnarray}}
\newcommand{\ba}{\begin{array}}
\newcommand{\ea}{\end{array}}

\newcommand{\bef}{\begin{figure}}
\newcommand{\eef}{\end{figure}}
\newcommand{\bce}{\begin{center}}
\newcommand{\ece}{\end{center}}

\def\la{\langle}
\def\ra{\rangle}

\newcommand{\gsim}{\mathrel{\hbox{\rlap{\lower.55ex \hbox {$\sim$}}
                   \kern-.3em \raise.4ex \hbox{$>$}}}}
\newcommand{\lsim}{\mathrel{\hbox{\rlap{\lower.55ex \hbox {$\sim$}}
                   \kern-.3em \raise.4ex \hbox{$<$}}}}

\begin{document}
\title*{Phases of QCD at High Baryon Density}
\toctitle{Phases of QCD at High Baryon Density}
%
%
\titlerunning{Phases of QCD at High Baryon Density}
%
\author{Thomas Sch\"afer\inst{1}\inst{2}
\and Edward Shuryak\inst{1}}
\authorrunning{T. Sch\"afer and E. Shuryak}
%
%
\institute{ Department of Physics and Astronomy, 
     State University of New York, 
     Stony Brook, NY 11794-3800 
\and 
Riken-BNL Research Center, Brookhaven National 
     Laboratory, Upton, NY 11973
}

\maketitle              

\begin{abstract}
 We review recent work on the phase structure of QCD 
at very high baryon density. We introduce the phenomenon
of color superconductivity and discuss how the quark
masses and chemical potentials determine the structure 
of the superfluid quark phase. We comment on the possibility 
of kaon condensation at very high baryon density and study 
the competition between superfluid, density wave, and chiral 
crystal phases at intermediate density. 
\end{abstract}

\section{Color Superconductivity}
\label{sec_intro}

  In the interior of compact stars matter is compressed
to densities several times larger than the density of 
ordinary matter. Unlike the situation in relativistic
heavy ion collisions, these conditions are maintained 
for essentially infinite periods of time. Also, compared
to QCD scales, matter inside a compact star is quite cold. 
At low density quarks are confined, chiral 
symmetry is broken, and baryonic matter is described in 
terms of neutrons and protons as well as their excitations.
At very large density, on the other hand, we expect that 
baryonic matter is described more effectively in terms
of quarks rather than hadrons. As we shall see, these
quarks can form new condensates and the phase structure
of dense quark matter is quite rich. 

  At very high density the natural degrees of freedom are
quark excitations and holes in the vicinity of the Fermi
surface. Since the Fermi momentum is large, asymptotic freedom
implies that the interaction between quasi-particles is weak.
In QCD, because of the presence of unscreened long range gauge 
forces, this is not quite true. Nevertheless, we believe 
that this fact does not essentially modify the argument. 
We know from the theory of superconductivity the Fermi 
surface is unstable in the presence of even an arbitrarily 
weak attractive interaction. At very large density, the 
attraction is provided by one-gluon exchange between quarks 
in a color anti-symmetric $\bar 3$ state. High density quark 
matter is therefore expected to be a color superconductor 
\cite{Frau_78,Barrois:1977xd,Bar_79,Bailin:1984bm}.

  Color superconductivity is described by a pair condensate
of the form
\beq
\label{csc}
\phi = \langle \psi^TC\Gamma_D\lambda_C\tau_F\psi\rangle.
\eeq
Here, $C$ is the charge conjugation matrix, and $\Gamma_D,
\lambda_C,\tau_F$ are Dirac, color, and flavor matrices. 
Except in the case of only two colors, the order parameter
cannot be a color singlet. Color superconductivity is 
therefore characterized by the breakdown of color gauge 
invariance. As usual, this statement has to be interpreted 
with care because local gauge invariance cannot really be 
broken. Nevertheless, we can study gauge invariant 
consequences of a quark pair condensate, in particular
the formation of a gap in the excitation spectrum.

 In addition to that, color superconductivity can lead to 
the breakdown of global symmetries. We shall see that in 
some cases there is a gauge invariant order parameter 
for the $U(1)$ of baryon number. This corresponds to true 
superfluidity and the appearance of a massless phonon. 
We shall also find that for $N_f>2$ color superconductivity 
leads to chiral symmetry breaking and that quark matter 
may support a kaon condensate. Finally, as we move to 
stronger coupling we find that other forms of order can 
compete with color superconductivity and that quark matter 
may exist in the form of chiral density waves or chiral 
crystals.

\section{Phase Structure in Weak Coupling}
\label{sec_phases}
\subsection{QCD with two flavors}
\label{sec_nf2}

  In this section we shall discuss how to use weak coupling
methods in order to explore the phases of dense quark matter.
We begin with what is usually considered to be the simplest 
case, quark matter with two degenerate flavors, up and down. 
Renormalization group arguments suggest 
\cite{Evans:1999ek,Evans:1999nf,Schafer:1999na}, and explicit
calculations show \cite{Brown:1999yd,Schafer:2000tw}, that
whenever possible quark pairs condense in an $s$-wave. This
means that the spin wave function of the pair is anti-symmetric. 
Since the color wave function is also anti-symmetric, the Pauli
principle requires the flavor wave function to be anti-symmetric,
too. This essentially determines the structure of the order
parameter \cite{Alford:1998zt,Rapp:1998zu}
\beq
\phi^a  = \langle \epsilon^{abc}\psi^b C\gamma_5 \tau_2\psi^c
 \rangle.
\eeq
This order parameter breaks the color $SU(3)\to SU(2)$ and
leads to a gap for up and down quarks with two out of the 
three colors. Chiral and isospin symmetry remain unbroken. 

  We can calculate the magnitude of the gap and the 
condensation energy using weak coupling methods. In weak
coupling the gap is determined by ladder diagrams with 
the one gluon exchange interaction. These diagrams can 
be summed using the gap equation 
\cite{Son:1999uk,Schafer:1999jg,Pisarski:2000tv,Hong:2000fh,Brown:1999aq}
\bea
\label{eliash}
\Delta(p_0) &=& \frac{g^2}{12\pi^2} \int dq_0\int d\cos\theta\,
 \left(\frac{\frac{3}{2}-\frac{1}{2}\cos\theta}
            {1-\cos\theta+G/(2\mu^2)}\right. \\
 & & \hspace{3cm}\left.    +\frac{\frac{1}{2}+\frac{1}{2}\cos\theta}
            {1-\cos\theta+F/(2\mu^2)} \right)
 \frac{\Delta(q_0)}{\sqrt{q_0^2+\Delta(q_0)^2}}. \nonumber
\eea
Here, $\Delta(p_0)$ is the frequency dependent gap, $g$ is the 
QCD coupling constant and $G$ and $F$ are the self energies of
magnetic and electric gluons. This gap equation is very similar
to the BCS gap equations that describe nuclear superfluids. 
The main difference is that because the gluon is massless, 
the gap equation contains a collinear divergence for $\cos
\theta\to 1$. In a dense medium the collinear divergence is 
regularized by the gluon self energy. For $\vec{q}\to 0$ 
and to leading order in perturbation theory we have
\beq
 F = 2m^2, \hspace{1cm}
 G = \frac{\pi}{2}m^2\frac{q_0}{|\vec{q}|},
\eeq
with $m^2=N_fg^2\mu^2/(4\pi^2)$. In the electric part,
$m_D^2=2m^2$ is the familiar Debye screening mass. In the 
magnetic part, there is no screening of static modes, 
but non-static modes are modes are dynamically screened
due to Landau damping.

 We can now perform the angular integral and find
\beq
\label{eliash_mel}
\Delta(p_0) = \frac{g^2}{18\pi^2} \int dq_0
 \log\left(\frac{b\mu}{|p_0-q_0|}\right)
    \frac{\Delta(q_0)}{\sqrt{q_0^2+\Delta(q_0)^2}},
\eeq
with $b=256\pi^4(2/N_f)^{5/2}g^{-5}$. This result shows
why it was important to keep the frequency dependence of the 
gap. Because the collinear divergence is regulated by
dynamic screening, the gap equation depends on $p_0$
even if the frequency is small. We can also see that
the gap scales as $\exp(-c/g)$. The collinear divergence 
leads to a gap equation with a double-log behavior. 
Qualitatively
\beq
\label{dlog}
 1 \sim \frac{g^2}{18\pi^2}
 \left[\log\left(\frac{\mu}{\Delta}\right)\right]^2,
\eeq
from which we conclude that $\Delta\sim\exp(-c/g)$. 
The approximation (\ref{dlog}) is not sufficiently
accurate to determine the correct value of the 
constant $c$. A more detailed analysis shows that
the gap on the Fermi surface is given by
\beq
\label{gap_oge}
\Delta_0 \simeq 512\pi^4(2/N_f)^{5/2}\mu g^{-5}
   \exp\left(-\frac{3\pi^2}{\sqrt{2}g}\right).
\eeq
We should emphasize that, strictly speaking, this result
contains only an estimate of the pre-exponent. It was 
recently argued that wave function renormalization
and quasi-particle damping give $O(1)$ contributions
to the pre-exponent which substantially reduce the 
gap \cite{Brown:1999aq}.

 For chemical potentials $\mu<1$ GeV, the coupling 
constant is not small and the applicability of perturbation
theory is in doubt. If we ignore this problem and extrapolate
the perturbative calculation to densities $\rho\simeq 5\rho_0$
we find gaps $\Delta\simeq 100$ MeV. This result may indeed 
be more reliable than the calculation on which it is based.
In particular, we note that similar results have been obtained
using realistic interactions which reproduce the chiral 
condensate at zero baryon density \cite{Alford:1998zt,Rapp:1998zu}.
 
 We can also determine the condensation energy. In weak coupling
the grand potential can be calculated from \cite{Bailin:1984bm}
\beq
\label{f_ng}
\Omega = \frac{1}{2}\int\frac{d^4q}{(2\pi)^4}
 \left\{-{\rm tr}\left[S(q)\Sigma(q)\right]
       +{\rm tr}\left[S_0^{-1}(q)S(q)\right]\right\},
\eeq
where $S(q)$ and $\Sigma(q)$ are the Nambu-Gorkov propagator 
and proper self energy. Using the propagator in the
superconducting phase we find
\beq
\epsilon = 4\frac{\mu^2}{4\pi^2} \Delta_0^2
 \log\left(\frac{\Delta_0}{\mu}\right),
\eeq
where the factor $4=2N_f$ comes from the number of 
condensed species. Using $\Delta_0\simeq 100$ MeV 
and $\mu\simeq 500$ MeV we find $\epsilon\simeq 
-50\,{\rm MeV}/{\rm fm}^3$. This shows that the 
condensation energy is only a small $O(\Delta^2/\mu^2)$
correction to the total energy density of the quark 
phase. We note that the result for the condensation 
energy agrees with BCS theory. The same is true for
the critical temperature which is given by $T_c=0.56
\Delta_0$. 
 
\subsection{QCD with three flavors: Color-Flavor-Locking}
\label{sec_cfl} 

 If quark matter is formed at densities several times 
nuclear matter density we expect the quark chemical 
potential to be larger than the strange quark mass.
We therefore have to determine the structure of the 
superfluid order parameter for three quark flavors.
We begin with the idealized situation of three 
degenerate flavors. From the arguments given in the 
last section we expect the order parameter to be 
color and flavor anti-symmetric matrix of the form
\beq
\label{order}
  \phi^{ab}_{ij}=
  \langle \psi^a_i C\gamma_5\psi^b_j\rangle.
\eeq
In order to determine the precise structure of this
matrix we have to extremize grand canonical potential.
We find \cite{Schafer:1999fe,Evans:1999at}
\beq
\label{cfl}
\Delta^{ab}_{ij} = 
 \Delta (\delta_i^a\delta_j^b-\delta_i^b\delta_j^a),
\eeq
which describes the color-flavor locked phase proposed in 
\cite{Alford:1999mk}. Both color and flavor symmetry are completely 
broken. There are eight combinations of color and flavor symmetries 
that generate unbroken global symmetries. The symmetry breaking
pattern is 
\beq
\label{sym_3}
SU(3)_L\times SU(3)_R\times U(1)_V\to SU(3)_V .
\eeq
This is exactly the same symmetry breaking that QCD exhibits 
at low density. The spectrum of excitations in the color-flavor-locked 
(CFL) phase also looks remarkably like the spectrum of QCD at low 
density \cite{Schafer:1999ef}. The excitations can be classified 
according to their quantum numbers under the unbroken $SU(3)$, and 
by their electric charge. The modified charge operator that generates 
a true symmetry of the CFL phase is given by a linear combination 
of the original charge operator $Q_{em}$ and the color hypercharge 
operator $Q={\rm diag}(-2/3,-2/3,1/3)$. Also, baryon number is only 
broken modulo 2/3, which means that one can still distinguish baryons 
from mesons. We find that the CFL phase contains an octet of Goldstone 
bosons associated with chiral symmetry breaking, an octet of vector 
mesons, an octet and a singlet of baryons, and a singlet Goldstone 
boson related to superfluidity. All of these states have integer 
charges.  

  With the exception of the $U(1)$ Goldstone boson, these states
exactly match the quantum numbers of the lowest lying multiplets
in QCD at low density. In addition to that, the presence of the 
$U(1)$ Goldstone boson can also be understood. The $U(1)$ order
parameter is $\langle (uds)(uds)\rangle$. This order parameter
has the quantum numbers of a $0^+$ $\Lambda\Lambda$ pair condensate.
In $N_f=3$ QCD, this is the most symmetric two nucleon channel, 
and a very likely candidate for superfluidity in nuclear matter
at low to moderate density. We conclude that in QCD with three
degenerate light flavors, there is no fundamental difference 
between the high and low density phases. This implies that a
low density hyper-nuclear phase and the high density quark phase
might be continuously connected, without an intervening phase
transition. 

\subsection{QCD with one flavor: Color-Spin-Locking}
\label{sec_csl}

 In this section we shall study superfluid quark matter with 
only one quark flavor. As we will discuss in more detail in 
the next section, our results are relevant to quark matter
with two or three flavors. This is the case when the Fermi 
surfaces of the individual flavors are too far apart in 
order to allow for pairing between different species. 

 If two quarks with the same flavor are in a color anti-symmetric 
wave function then there combined spin and spatial wave function 
cannot be antisymmetric. This means that pairing between 
identical flavors has to involve total angular momentum one 
or greater. The simplest order parameters are of the form 
\beq
\label{spin1}
\Phi_1=\langle \epsilon^{3ab}\psi^a C\vec\gamma \psi^b\rangle,
\hspace{1cm}
\Phi_2=\langle \epsilon^{3ab}\psi^a C\hat{\vec{q}} \psi^b\rangle.
\eeq
The corresponding gaps can be determined using the methods 
introduced in section \ref{sec_nf2}. We find $\Delta(
\Phi_{1,2})=\exp(-3c_{1,2})\Delta_0$ with $c_1=-1.5$ and 
$c_2=-2$ \cite{Brown:1999yd,Schafer:2000tw}. While the natural 
scale of the s-wave gap is $\Delta=100$ MeV, the p-wave gap is 
expected to be less than 1 MeV.
 
 The spin one order parameter (\ref{spin1}) is a color-spin
matrix. This opens the possibility that color and spin degrees
become entangled, similar to the color-flavor locked phase 
(\ref{cfl}) or the B-phase of liquid $^3$He. The corresponding
order parameter is
\beq
\label{csl}
 \Phi_{CSL} = \delta^a_i \langle\epsilon^{abc}\psi^b C 
 \left(\cos(\beta)\hat{q}_i+\sin(\beta)\gamma_i \right)
 \psi^c\rangle,
\eeq
where the angle $\beta$ determines the mixing between the 
two types of condensates shown in (\ref{spin1}). In reference
\cite{Schafer:2000tw} we showed that the color-spin locked
phase (\ref{csl}) is favored over the ``polar'' phase
(\ref{spin1}). 

 In the color-spin locked phase color and rotational invariance
are broken, but a diagonal $SO(3)$ survives. As a consequence,
the gap is isotropic. There are no gapless modes except if 
$\beta$ takes on special values. The parameter $\beta$ is
sensitive to the quark mass and to higher order corrections.
In the non-relativistic limit we find $\beta=\pi/4$ whereas 
in the ultra-relativistic limit $\beta=\pi/2$.

\section{The Role of the Strange Quark Mass and the 
Electron Chemical Potential} 
\label{sec_unlock}

  So far, we have only considered the case of two or three
degenerate quark flavors. In the real world, the strange quark 
is significantly heavier than the up and down quarks. Also, in 
the interior of a neutron star, electrons are present and the
chemical potentials for up and down quarks are not equal.

  The role of the strange quark mass in the high density phase 
was studied in \cite{Alford:1999pa,Schafer:1999pb}. The main effect 
is a purely kinematic phenomenon that is easily explained. The Fermi 
surface for the strange quarks is shifted by $\delta p_F=\mu-(\mu^2
-m_s^2)^{1/2}\simeq m_s^2(2\mu)$ with respect to the Fermi surface 
of the light quarks. The condensate involves pairing between quarks 
of different flavors at opposite points on the Fermi surface. But 
if the Fermi surfaces are shifted, then the pairs do not have total 
momentum zero, and they cannot mix with pairs at others points on 
the Fermi surface. If the system is superfluid then the Fermi surface 
is smeared out over a range $\Delta$. This means that pairing
between strange and light quarks can take place as long as the 
mismatch between the Fermi momenta is smaller than the gap, 
\beq
\label{m_s_crit}
 \Delta > \frac{m_s^2}{2\mu}.
\eeq
This conclusion is supported by a more detailed analysis
\cite{Alford:1999pa,Schafer:1999pb}. Since flavor symmetry 
is broken, we allow the $\langle ud\rangle$ and $\langle us\rangle =
\langle ds\rangle$ components of the CFL condensate to be 
different. The $N_f=2$ phase corresponds to $\langle us
\rangle = \langle ds\rangle =0$. We find that there is a 
first order phase transition from the CFL to the $N_f=2$ 
phase, and that the critical strange quark mass is in rough 
agreement with the estimate (\ref{m_s_crit}). 

 For densities $\rho\simeq (5-10)\rho_0$ the critical strange
quark mass is close to physical mass of the strange quark. It
is therefore hard to predict with certainty whether superconducting
strange quark matter in the interior of a neutron star is in the
color-flavor locked phase. Neutron star observations may help 
to answer the question. In the CFL phase all quarks have large
gaps, whereas in the unlocked phase up and down quarks of the
first two colors have large gaps, strange quarks have a small
gap, and the up and down quarks of the remaining color have 
tiny gaps, or may not be gapped at all. 

 The effects of a non-zero electron chemical potential was 
studied in \cite{Bedaque:1999nu}. If the electron chemical
potential exceeds the gap in the up-down sector then up 
and down quarks cannot pair. In this case, the up and
down quarks pair separately, and with much smaller gaps, 
in the one flavor phase discussed in section \ref{csl}. 
Between the phases with $\langle ud\rangle$ and $\langle
uu\rangle,\,\langle dd\rangle$ pairing there is a small
window of electron chemical potentials where inhomogeneous
superconductivity takes place \cite{Alford:2000ze}.

\section{Kaon Condensation}
\label{sec_kaon}
 
 The low energy properties of dense quark matter are 
determined by collective modes. In the color-flavor
locked phase these modes are the phonon and the 
pseudoscalar Goldstone bosons, the pions, kaons and
etas. Some time ago, it was suggested that pions 
\cite{Migdal:1973,Sawyer:1972,Scalapino:1972} or 
kaons \cite{Kaplan:1986,Brown:1987} might form Bose 
condensates in dense baryonic matter. The proposed
critical densities were close to nuclear matter 
density in the case of pion condensation, and 
several times nuclear matter density in the case 
of kaon condensation. Since mesonic modes persist
in the color-flavor-locked phase of quark matter, we
can now revisit the issue of Goldstone boson 
condensation \cite{Schafer:2000ew}. In particular,
we shall be able to use rigorous weak coupling methods
in order to address the possibility of Bose condensation
in dense matter.  
 
 Our starting point is the effective lagrangian for the 
pseudoscalar Goldstone in dense matter \cite{Casalbuoni:1999,Son:1999}
\beq
\label{leff}
{\cal L}_{eff} = \frac{f_\pi^2}{4} {\rm Tr}\left[
 \partial_0\Sigma\partial_0\Sigma^\dagger + v_\pi^2
 \partial_i\Sigma\partial_i\Sigma^\dagger \right]
 - c\left[\det(M){\rm Tr}(M^{-1}\Sigma) + h.c.\right].
\eeq
Here, $\Sigma\in SU(3)$ is the Goldstone boson field, 
$v_\pi$ is the velocity of the Goldstone modes and 
$M={\rm diag}(m_u,m_d,m_s)$ is the quark mass matrix. 
The effective description is valid for energies and
momenta below the scale set by the gap, $\omega,q\ll
\Delta$. The low energy constants can be determined
in weak coupling perturbation theory. The result is
$v_\pi^2=1/3$ and 
\cite{Son:1999,Rho:1999,Hong:1999,Manuel:2000,Beane:2000}
\bea
f_\pi^2 &=& \frac{21-8\log(2)}{18} \frac{\mu^2}{2\pi^2}, \\
 c &=& \frac{3\Delta^2}{2\pi^2}\cdot\frac{2}{f_\pi^2}.
\eea
We can now determine the masses of the Goldstone 
bosons
\beq 
\label{mgb}
 m_{\pi^\pm} = c(m_u+m_d)m_s, \hspace{1cm}
 m_{K_\pm}   = cm_d (m_u+m_s).
\eeq
This result shows that the kaon is {\em lighter} than 
the pion. This can be understood from the fact that,
at high density, it is more appropriate to think
of the interpolating field $\Sigma$ as
\beq
\label{sigma}
 \Sigma_{ij} \sim \epsilon_{ikl}\epsilon_{jmn}
 \bar\psi_{L,k}\bar\psi_{L,l}\psi_{R,m}\psi_{R,n}
\eeq
rather than the more familiar $\Sigma_{ij}\sim
\bar\psi_{L,i}\psi_{R,j}$ \cite{Casalbuoni:1999}.
Using (\ref{sigma}) we observe that the negative pion 
field has the flavor structure $\bar{d}\bar{s}us$ and 
therefore has mass proportional to $(m_u+m_d)m_s$
\cite{Son:1999}. Putting in numerical values we
find that the kaon mass is very small, $m_{K^{-}} 
\simeq 5$ MeV at $\mu=500$ MeV and $m_{K^{-}}
\simeq 1$ MeV at $\mu=1000$ MeV. 

  There are two reasons why the pseudoscalar Goldstone
bosons are anomalously light. First of all, the Goldstone 
boson masses in the color-flavor-locked phase are proportional 
to the quark masses squared rather than linear in the 
quark mass, as they are at zero density. This is due to an 
approximate $Z_2$ chiral symmetry in the color-flavor-locked 
phase \cite{Alford:1999mk}. In addition to that, the Goldstone 
boson masses are suppressed by a factor $\Delta/\mu$. This is
a consequence of the fact that the Goldstone modes are collective 
excitations of particles and holes near the Fermi surface, 
whereas the quark mass term connects particles and anti-particles, 
far away from the Fermi surface. 

\begin{figure}[t]
\begin{center}
\leavevmode
\vspace{1cm}
\epsfig{file=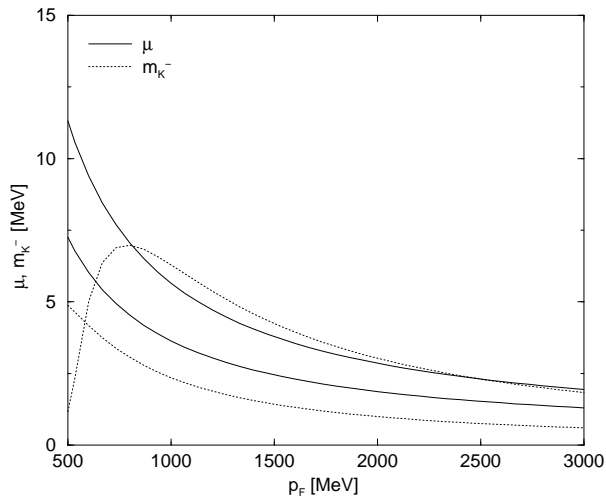,width=8cm}
\end{center}
\caption{\label{fig_kaon}
Electron chemical potential (solid lines) and kaon mass
(dashed lines) in the color-flavor-locked quark phase. The two 
curves for both quantities represent a simple estimate of the 
uncertainties due to the value of the strange quark mass and
the scale setting procedure.}
\end{figure}

  The fact that the meson spectrum is inverted, and that the kaon 
mass is exceptionally small opens the possibility that in dense quark 
matter electrons decay into kaons, and a kaon condensate is formed. 
Consider a kaon condensate $\langle K^-\rangle = v_Ke^{-i\mu t}$ 
where $\mu$ is the chemical potential for negative charge. The 
thermodynamic potential ${\cal H}-\mu Q$ for this state is given 
by
\bea
\epsilon(\rho_q,x,y,\mu) &=&  \frac{3}{4\pi^2}
\pi^{8/3}\rho_q^{4/3} \left\{ x^{4/3} + y^{4/3}
 + (1-x-y)^{4/3} \right. \nonumber \\
 & & \hspace{0.5cm}\left.\mbox{}+ \pi^{-4/3}\rho_q^{-2/3}m_s^2
  (1-x-y)^{2/3} \right\} \nonumber \\
 & & \hspace{0.5cm}\mbox{}- \left(\mu^2-m_K^2\right)v_K^2 +O(v_K^3)
 + \mu \rho_q x - \frac{1}{12\pi^2}\mu^4
\label{eps}
\eea
Here, $\rho_q=3\rho_B$ is the quark density, and $x=\rho_u/
\rho_q$ and $y=\rho_d/\rho_q$ are the up and down quark 
fractions. For simplicity, we have dropped higher order terms 
in the strange quark mass and neglected the electron mass. 
In order to determine the ground state we have to make 
(\ref{eps}) stationary with respect to $x,y,\mu$ and $v_K$. 
Minimization with respect to $x$ and $y$ enforces $\beta$ 
equilibrium, while minimization with respect to $\mu$ ensures 
charge neutrality. Below the onset for kaon condensation 
we have $v_K=0$ and there is no kaon contribution to the
charge density. Neglecting $m_e$ and higher order corrections
in $m_s$ we find 
\beq
 \mu\simeq \frac{m_s^2}{4p_F} 
  = \frac{\pi^{2/3}m_s^2}{4\rho_B^{1/3}}.
\eeq
In the absence of kaon condensation, the electron chemical
potential will level off at the value of the electron mass
for very high baryon density. The onset of kaon condensation
is determined by the condition $\mu=m_K$. At this point it 
becomes favorable to convert electrons into negatively 
charged kaons. 

 Results for the electron chemical potential and the
kaon mass as a function of the light quark Fermi
momentum are shown in Fig. \ref{fig_kaon}. In order to assess
some of the uncertainties we have varied the quark 
masses in the range $m_u=(3-5)$ MeV, $m_d=(6-8)$ MeV,
and $m_s=(120-150)$ MeV. We have used the one loop
result for the running coupling constant at two 
different scales $q=\mu$ and $q=\mu/2$. The scale
parameter was set to $\Lambda_{QCD}=238$ MeV,
corresponding to $\alpha_s(m_\tau)=0.35$. An important 
constraint is provided by the condition $m_s<\sqrt{2p_F\Delta}$ 
discussed in the previous section. We have checked that this 
condition is always satisfied for $p_F>500$ MeV. Figure 
\ref{fig_kaon} shows that there is significant uncertainty 
in the relative magnitude of the chemical potential and the 
kaon mass. Nevertheless, the band of kaon mass predictions 
lies systematically below the predicted chemical potentials. 
We therefore conclude that kaon condensation appears likely even 
for moderate Fermi momenta $p_F\simeq 500$ MeV. For
very large baryon density $\mu\to m_e$ while $m_K\to 0$
and kaon condensation seems inevitable.

\section{Chiral Waves and Chiral Crystals}
\label{sec_crystal}

 It is very important for the structure of compact stars to 
determine whether the quark matter core is in liquid or solid 
form. In the previous sections we discussed the case of weak 
coupling. In this case, particle-particle pairing is the only
instability that needs to be considered. Nevertheless, in strong 
coupling, or if superconductivity is suppressed, other forms 
of pairing may take place. Obvious candidates are the formation of 
larger clusters or particle-hole pairing.

 Particle-hole pairing is characterized by an order parameter of
the form
\beq 
\langle\bar\psi(x)\psi(y) \rangle = 
 \exp(i\vec{Q}\cdot(\vec{x}+\vec{y}))\Sigma(x-y),
\eeq
where $\vec{Q}$ is an arbitrary vector. This state describes a 
chiral density wave. It was first suggested in \cite{Deryagin:1992} 
as the ground state of QCD at large chemical potential and large $N_c$. 
This suggestion was based on the fact that particle-particle pairing, 
and superconductivity, is suppressed for large $N_c$ whereas 
particle-hole pairing is not. Particle-hole pairing, on the 
other hand, uses only a small part of the Fermi surface and
does not take place in weak coupling. In the case of the 
one-gluon exchange interaction these issues were studied in
\cite{Shuster:1999}. The main conclusion is that, in weak
coupling, the chiral density wave instability requires very
large $N_c\gg 3$.

 At moderate densities, and using realistic interactions, this
is not necessarily the case. In particular, we know that at zero
density the particle-anti-particle interaction is stronger, by
a factor $N_c-2$, than the particle-particle interaction. In 
a Nambu-Jona-Lasinio type description this interaction 
exceeds the critical value required for chiral symmetry breaking
to take place. For this reason we have recently studied the 
competition between the particle-particle and particle-hole 
instabilities using non-perturbative, instanton generated, forces 
\cite{Rapp:2000}. Our results are not only relevant to 
flavor symmetric quark matter at moderate densities, but 
also for the important case when there is a substantial 
difference between the chemical potentials for up and down
quarks. As discussed in section \ref{sec_unlock} this disfavors
$ud$-pairing, but it does not inhibit $uu^{-1}$ and $dd^{-1}$
particle-hole pairing. 

  Our results show that at low density the chiral density 
wave state is practically degenerate with the BCS solution. 
Given the uncertainties that affect the calculation this 
implies that both states have to be considered as realistic
possibilities for the behavior of quark matter near the 
phase transition.

\subsection{BCS Pairing}

  In order to study competing instabilities we use the standard 
Nambu-Gorkov formalism, in which the propagator is written as a 
matrix in the space of all possible pair condensates. The BCS
channel is described by the $2\times 2$ matrix
\beq
\hat{G}_{BCS}
=\left( \ba{cc}
\la c_{k\uparrow} ~ c^\dagger_{k\uparrow} \ra  &
\la c_{k\uparrow} ~ c_{-k\downarrow}      \ra \\
\la c^\dagger_{-k\downarrow} ~ c^\dagger_{k\uparrow} \ra  &
\la c^\dagger_{-k\downarrow} ~ c_{-k\downarrow} \ra
               \ea \right) \
\equiv\left( \ba{cc}
G(k_0,\vec k,\Delta)  & \bar{F}(k_0,\vec k,\Delta)
\\
F(k_0,\vec k,\Delta) & \bar{G}(k_0,-\vec k,\Delta)
               \ea \right) \ .
\label{G_bcs}
\eeq
The propagator has the form 
\beq
\hat{G}_{BCS}=\frac{1}{G_0^{-1} \bar{G}_0^{-1} -\Delta\bar{\Delta}}
\left( \ba{cc}
\bar{G}_0^{-1}  &  -{\Delta} \\
  -\bar{\Delta} & G_0^{-1}
               \ea \right) \ ,
\label{sol_bcs}
\eeq
where
\beq
G_0 = \frac{1}{k_0-\epsilon_k+i\delta_{\epsilon_k}},
\hspace{0.5cm}
\bar{G}_0 = \frac{1}{k_0+\epsilon_k+i\delta_{\epsilon_k}}
\eeq
are the free particle propagator and its conjugate at finite
chemical potential. Here, $\epsilon_k=\omega_k-\mu_q$ and 
$\delta_{\epsilon_k}={\rm sgn}(\epsilon_k)\delta$ determines
the pole position. From this equation we can read off the 
diagonal and off-diagonal components of the Gorkov propagator.
The off-diagonal, anomalous, propagator is
\beq
F(k_0,\vec k,\Delta)=\frac{-\Delta}{(k_0-\epsilon_k+i\delta_{\epsilon_k})
(k_0+\epsilon_k+i\delta_{\epsilon_k})-\Delta^2} \ .
\eeq
The anomalous self energy $\Delta$ is determined by the gap equation 
\beq
\Delta= (-i) \ \alpha_{pp} \int\frac{d^4p}{(2\pi)^4} \ 
 F(p_0,\vec p,\Delta).
\label{gap_bcs}
\eeq
Here $\alpha_{pp}$ is the effective coupling in the particle-particle 
channel.

\subsection{Chiral Density Wave}

 Using the same formalism we can also address pairing in the 
particle-hole channel at finite total pair momentum $Q$. In the 
mean-field approximation, the full Greens function in the presence
of a single density wave takes the form
\bea
\hat{G}_{Ovh}
&=& \left( \ba{cc}
\la c_{k\uparrow} ~ c^\dagger_{k\uparrow} \ra  &
\la c_{k\uparrow} ~ c^\dagger_{k+Q\downarrow}      \ra \\
\la c_{k+Q\downarrow} ~ c^\dagger_{k\uparrow} \ra  &
\la c_{k+Q\downarrow} ~ c^\dagger_{k+Q\downarrow} \ra
               \ea \right) 
 \equiv \left( \ba{cc}
G(k_0,\vec k,\vec Q,\sigma)  & \bar{S}(k_0,\vec k,\vec Q,\sigma)
\\
S(k_0,\vec k,\vec Q,\sigma) & G(k_0,\vec k+\vec Q,\vec Q,\sigma)
               \ea \right) \nonumber \\
 &=& \left[ \hat{G}_0^{-1}-\hat{\sigma}\right]^{-1}
\eea
Again, we can read off the anomalous part of the Greens function
\beq
S(k_0,\vec k,\vec Q,\sigma)=\frac{-\sigma}
{(k_0-\epsilon_k+i\delta_{\epsilon_k}) 
(k_0-\epsilon_{k+Q}+i\delta_{\epsilon_{k+Q}})-\sigma^2},
\label{prop_ovh}
\eeq
and the anomalous self energy $\sigma$ is determined by a 
self-consistency, or gap, equation
\beq
\sigma=(-i) \alpha_{ph}\int\frac{d^4p}{(2\pi)^4} \ 
S(p_0,\vec p,\vec Q,\sigma;\mu_q) \ .
\label{gap_ovh}
\eeq
Notice that the energy contour integration receives non-vanishing
contributions only if
\beq
\epsilon_p \ \epsilon_{p+Q} - \sigma^2< 0 \ ,
\eeq
which ensures that the two poles in $p_0$ are in different (upper/lower)
half-planes. This means that one particle (above the Fermi surface) and 
one hole (below the Fermi surface) participate in the interaction. 

 The formation of a condensate carrying nonzero total momentum 
$\vec{Q}$ is associated with nontrivial spatial structures. In 
the simplest case of particle-hole pairs with total momentum $Q$ 
this is a density wave of wave length $\lambda=2\pi/Q$. In three 
dimensions, however, we can have several density waves characterized 
by different momenta $\vec{Q}$. In this case, the resulting spatial
structure is a crystal. In general, the $p$-$h$ pairing gap can 
be written as
\beq
\sigma(\vec{r})=\sum\limits_j \sum\limits_{n=-\infty}^{+\infty}
\sigma_{j,n} e^{in\vec{Q}_j\cdot \vec{r}} \ ,
\eeq
where the $\vec{Q}_j$ correspond to the (finite) number of fundamental 
waves, and the summation over $|n|>1$ accounts for higher harmonics 
in the Fourier series. The matrix propagator formalism allows for the 
treatment of more than one density wave through a straightforward 
expansion of the basis states according to
\beq
\hat{G}=\left( \ba{cccc}
\la c_{k\uparrow} ~ c^\dagger_{k\uparrow} \ra        &
\la c_{k\uparrow} ~ c^\dagger_{k+Q_x\downarrow} \ra  &
\la c_{k\uparrow} ~ c^\dagger_{k+Q_y\downarrow} \ra  &
 \cdots                                              \\
\la c_{k+Q_x\downarrow} ~ c^\dagger_{k\uparrow} \ra        &
\la c_{k+Q_x\downarrow} ~ c^\dagger_{k+Q_x\downarrow} \ra  &
\la c_{k+Q_x\downarrow} ~ c^\dagger_{k+Q_y\downarrow} \ra  &
 \cdots                                              \\
\la c_{k+Q_y\downarrow} ~ c^\dagger_{k\uparrow} \ra        &
\la c_{k+Q_y\downarrow} ~ c^\dagger_{k+Q_x\downarrow} \ra  &
\la c_{k+Q_y\downarrow} ~ c^\dagger_{k+Q_y\downarrow} \ra  &
 \cdots                                              \\
\vdots & \vdots & \vdots &\ddots  \\
\ea
\right)  \ .
\label{G_ovh}
\eeq
The possibility of simultaneous BCS pairing can be incorporated
by extending the Gorkov propagator to include both particle-hole
and particle-particle components. In the following we will consider 
up to $n_w=6$ waves in three orthogonal directions with $Q_x=Q_y=
Q_z$ and $n=\pm 1$, characterizing a cubic crystal.

\begin{figure}[t]
\vspace{-0.7cm}
\bce
\epsfig{file=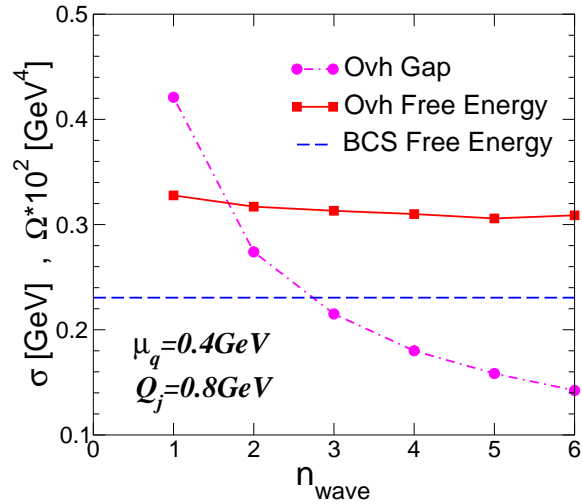,width=8cm,angle=-90}
\ece
\vspace{-0.3cm}
\caption{Dependence of the free energy (upper full line) and 
$p$-$h$ pairing gap (dashed-dotted) on the number of waves 
('patches') with fixed magnitude of the three-momentum 
$|\vec Q_j|=0.8$~GeV. The full line shows the value of the 
BCS ground state free energy. The results correspond to an
instanton calculation with $\mu_q=0.4$~GeV and $N/V=1$~fm$^{-4}$.}
\label{fig_Ndep}
\end{figure}

 Note that in the propagators $G_0$ we do not include the contribution
of anti-particles. This should be a reasonable approximation in the
quark matter phase at sufficiently large $\mu_q$, when the standard
particle-anti-particle chiral condensate has disappeared. At the same 
time, since our analysis is based on non-perturbative forces, the
range of applicability is limited from above. Taken together, we estimate 
the range of validity for our calculations to be roughly given by
0.4~GeV~$\gsim \mu_q \gsim$~0.6~GeV. This coincides with
the regime where, for the physical current strange quark mass of
$m_s\simeq 0.14$~GeV, the two-flavor superconductor might prevail
over the color-flavor locked (CFL) state so that our restriction
to $N_f=2$ is supported.

 Solutions of the gap equations correspond to extrema (minima) in the
energy density with respect to the gap $\sigma$. However, solutions
may exist for several values of the wave vector $Q$. To determine the
minimum in this quantity, one has to take into account the explicit form
of the free energy density. In the mean-field approximation,
\bea
V_3\,\Omega (\mu_q, Q, \sigma) = 
\int\, d^3x \,\left(  \frac{\sigma^2 (x)}{2\lambda} +
\left\langle q^\dagger \, ( i{\alpha}\cdot\nabla -2\sigma (x)\, q) 
 \right\rangle\right)\,\,,
\eea
where $V_3$ is the 3-volume. The first contribution removes the double 
counting from the fermionic contribution in the mean-field treatment.

\begin{figure}[t]
\vspace{-0.7cm}
\bce
\epsfig{file=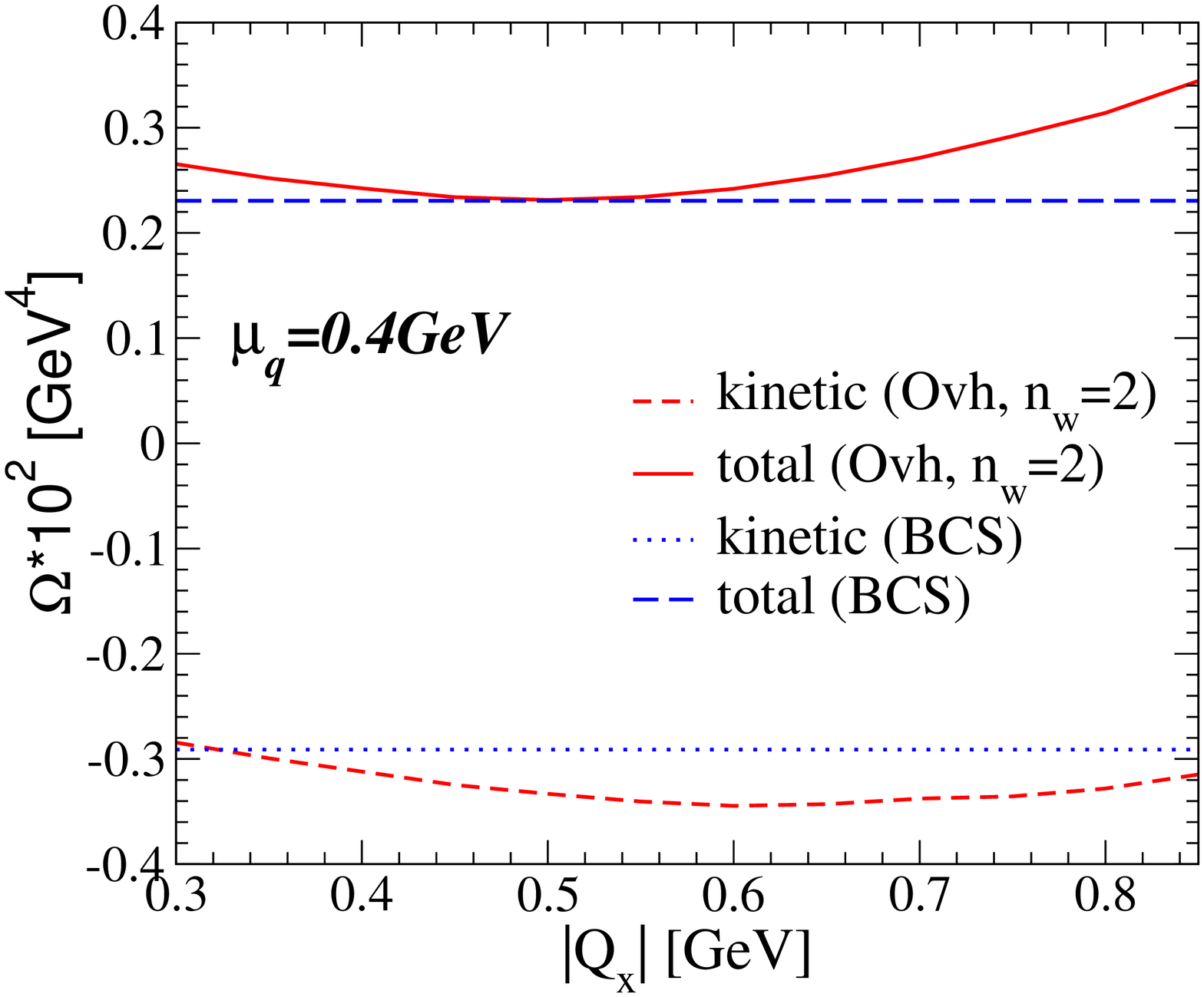,width=5cm,angle=-90}
\hspace{-1.cm}
\epsfig{file=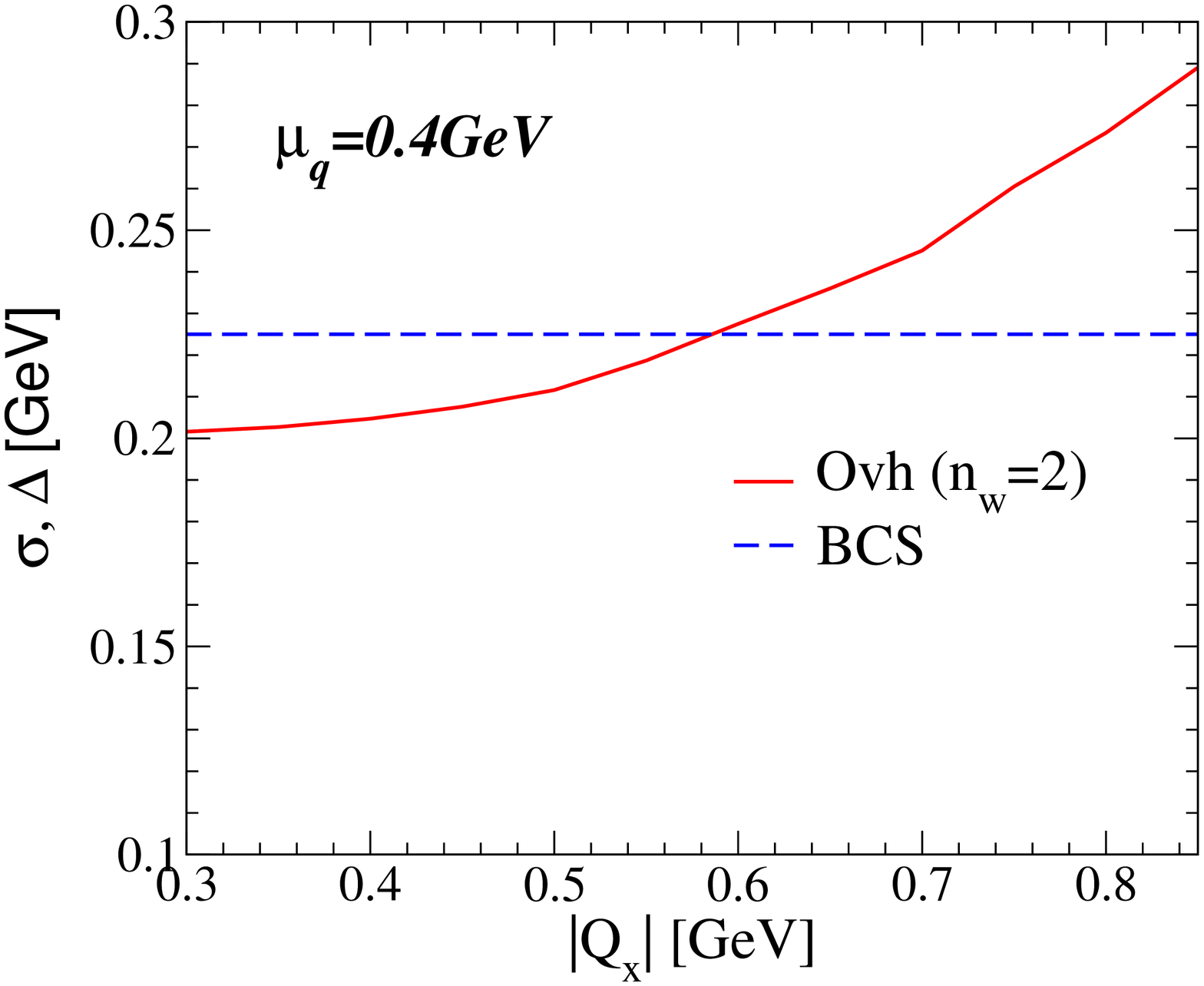,width=5cm,angle=-90}
\ece
\vspace{-0.3cm}
\caption{Left panel: wave-vector dependence of the density wave free energy 
for one standing wave  (full line: $\Omega_{tot}^{Ovh}$, short-dashed line: 
$\Omega_{kin}^{Ovh}$) in comparison to the BCS solution (long-dashed line: 
$\Omega_{tot}^{BCS}$, dotted line: $\Omega_{kin}^{BCS}$) at $\mu_q=0.4$GeV.
Right panel: wave-vector dependence of the density wave pairing gap (full 
line) compared to the BCS gap (long-dashed line).}
\label{fig_Qdep}
\end{figure}

 We have studied the coupled gap equations numerically. We do not
find any solutions with simultaneous particle-particle and
particle-hole condensates. This reduces the problem to the
question whether the BCS or the density wave state is 
thermodynamically favored. The BCS solution $\Delta=0.225$~GeV
is unique and has free energy of $\Omega_{BCS}(\mu_q=0.4~{\rm GeV})
=2.3\cdot 10^{-3}$~GeV$^4$. Here, we have neglected an irrelevant
overall constant that does not affect the comparison with the
density wave state.

 The situation is more complicated in the case of particle-hole
pairing. Let us start with the 'canonical' case where the momentum 
of the chiral density wave is fixed at twice the Fermi momentum, 
$Q=2p_F$. In fig.~\ref{fig_Ndep} the resulting minimized free energy 
is displayed as a function of the number of included waves. The 
density wave solutions are not far above the BCS groundstate, with 
a slight energy gain for an increased number of waves. 

 However, one can further economize the energy of the chiral density 
wave state by exploiting the freedom associated with the wave vector $Q$ 
(or, equivalently, the periodicity of the lattice). For $Q>2p_F$ the free 
energy rapidly increases. On the other hand, for $Q<2p_F$ more favorable 
configurations are found. To correctly assess them one has to include 
the waves in pairs $|k\pm Q_j|$ of standing waves ($n_w=2,4,6,\dots$)
to ensure that the occupied states in the Fermi sea are saturated
within the first Brillouin Zone. The lowest-lying state we could find 
at $\mu_q=0.4$~GeV occurs for one standing wave with $Q_{min}\simeq 
0.5$~GeV and $\sigma\simeq 0.21$~GeV with a free energy $\Omega\simeq 
2.3\cdot 10^{-3}$~GeV$^4$, practically degenerate with the BCS solution. 
This density wave has a wavelength $\lambda\simeq 2.5$~fm. The minimum 
in the wave vector is in fact rather shallow, as seen from the explicit 
momentum dependence of the free energy displayed in fig.~\ref{fig_Qdep}.

\section{Acknowledgements}

 We would like to thank our collaborators R.~Rapp, M.~Velkovsky,
F.~Wilzcek and I.~Zahed.


%

\end{document}